\documentclass[article,preprint,amsmath,amssymb,amsfonts,superscriptaddress]{revtex4-2}
\usepackage{times,graphicx,caption,epstopdf,dcolumn,bm,color,braket,multirow,xfrac,etoolbox}
\usepackage{hyperref} 
\usepackage{blindtext}
\usepackage[T1]{fontenc}
\usepackage[newcommands]{ragged2e}
\usepackage{amsmath}

\renewcommand{\figurename}{\textbf{{Fig.}}}
\renewcommand{\thefigure}{\textbf{\arabic{figure}}}
\DeclareCaptionLabelSeparator{bar}{ | }
\usepackage[T1]{fontenc}
\usepackage[font=small,labelfont=bf,justification=justified,format=plain]{caption} 
\setcitestyle{super} 

\usepackage{xr-hyper}
\makeatletter
\newcommand*{\addFileDependency}[1]{
\typeout{(#1)}
\@addtofilelist{#1}
\IfFileExists{#1}{}{\typeout{No file #1.}}
}\makeatother
\newcommand*{\myexternaldocument}[1]{
\externaldocument{#1}%
\addFileDependency{#1.tex}%
\addFileDependency{#1.aux}%
}
\myexternaldocument{SM} 

\begin{document}

\title{Origin of Distinct Insulating Domains in the Layered Charge Density Wave Material 1\textit{T}-TaS${_2}$}

\author{Hyungryul Yang}
\altaffiliation{\texorpdfstring{These authors contributed equally to this work}{}}
\affiliation{Department of Physics, Yonsei University, Seoul 03722, Korea}
\author{Byeongin Lee}
\altaffiliation{\texorpdfstring{These authors contributed equally to this work}{}}
\affiliation{Department of Physics, Yonsei University, Seoul 03722, Korea}
\author{Junho Bang}
\affiliation{Department of Physics, Yonsei University, Seoul 03722, Korea}
\author{Sunghun Kim}
\affiliation{Research Institute for Basic Science, Ajou University, Suwon 16499, Korea}
\author{Dirk Wulferding}
\affiliation{Center for Correlated Electron Systems, Institute for Basic Science, Seoul 08826, Korea}
\author{Sung-Hoon Lee} \email{lsh@khu.ac.kr}
\affiliation{Department of Applied Physics, Kyung Hee University, Yongin 17104, Korea}
\author{Doohee Cho} \email{dooheecho@yonsei.ac.kr}
\affiliation{Department of Physics, Yonsei University, Seoul 03722, Korea}

\maketitle

{\bf Vertical charge order shapes the electronic properties in layered charge density wave (CDW) materials. Various stacking orders inevitably create nanoscale domains with distinct electronic structures inaccessible to bulk probes. Here, the stacking characteristics of bulk 1T-TaS2 are analyzed using scanning tunneling spectroscopy (STS) and density functional theory (DFT) calculations. It is observed that Mott-insulating domains undergo a transition to band-insulating domains restoring vertical dimerization of the CDWs. Furthermore, STS measurements covering a wide terrace reveal two distinct band insulating domains differentiated by band edge broadening. These DFT calculations reveal that the Mott insulating layers preferably reside on the subsurface, forming broader band edges in the neighboring band insulating layers. Ultimately, buried Mott insulating layers believed to harbor the quantum spin liquid phase are identified. These results resolve persistent issues regarding vertical charge order in 1T-TaS2, providing a new perspective for investigating emergent quantum phenomena in layered CDW materials.}

\bigskip
\noindent\textit{\bf Introduction} 

Layered materials exhibit unique electronic properties governed by the atomic and electronic characteristics of their constituent building blocks, as well as their interlayer couplings, including Coulomb interactions, orbital hybridization, charge transfer, and van der Waals forces~\cite{alden2013strain,ritschel2015orbital,vavno2021artificial}. Given the susceptibility of these interlayer interactions to various stimuli, such as atomic defects~\cite{zhao2020engineering} and hydrostatic pressure~\cite{ritschel2013pressure}, distinct domains with various stacking configurations naturally coexist in layered materials. To understand the electronic properties of these diverse structures and the emergence of exotic quantum phenomena at their interfaces, it is crucial to systematically investigate domain configurations and their electronic structures.

1\textit{T}-TaS${_2}$ is an extensively studied two-dimensional correlated system, renowned for its rich phase diagram~\cite{ritschel2013pressure, fazekas1979electrical, sipos2008mott,  stojchevska2014ultrafast, yu2015gate, gerasimenko2019quantum, ravnik2021time, liu2023electrical, jarc2023cavity}. The atomic layer is composed of a plane of Ta atoms encapsulated by S atoms with an octahedral coordination. Upon cooling, it undergoes a metal-insulator transition with a charge density wave (CDW) reconstruction~\cite{wilson1975charge}. The CDW phase is accompanied by a $\sqrt{13} \times \sqrt{13}$ periodic lattice distortion, comprising clusters that resemble the Star of David (SD). Each SD contains 13 Ta atoms, with 12 forming pairs to create a CDW band gap, while the unpaired electron of the central Ta atom contributes to a narrow metallic band~\cite{fazekas1979electrical}. Due to strong on-site Coulomb interactions, intralayer electron hopping is strongly suppressed at low temperatures. Since the localized spins of the correlated insulator are frustrated by the triangular lattice, this material is an ideal platform to realize the quantum spin liquid phase (QSL)\cite{anderson1973resonating, law20171t}.

In bulk 1\textit{T}-TaS${_2}$, the electronic structure is significantly influenced by the vertical stacking order~\cite{wang2020band}, resulting in the ground state with a dimerized configuration. Interlayer electron hopping promotes a dispersive metallic band along the stacking direction~\cite{darancet2014three}.  However, at low temperatures, it exhibits insulating electronic properties with commensurate CDWs. Theoretical calculations associate this with the formation of a dimerized insulating layer where the SD is vertically aligned without a lateral shift~\cite{lee2019origin, petocchi2022mott}. This dimerized configuration is energetically stable because of the energy gain from bonding-antibonding hybridization of the unpaired electrons; otherwise, it would be prohibited due to Coulomb repulsion. Dimerized band insulating layers are stacked with a lateral shift in the bulk. This is similar to the Peierls' CDW metal-insulator transition in one-dimensional half-filled metallic systems~\cite{gruner1988dynamics}.

Cleaving this layered material produces two distinct insulating surfaces: Type-I with a dimerized layer termination and Type-II with an undimerized layer termination. The two distinct surface terminations can be distinguished by their energy band gaps~\cite{petocchi2022mott, butler2020mottness, lee2023charge}, with Type-I exhibiting a larger band gap compared to Type-II. They also exhibit distinct spectral responses to electron doping, featuring a rigid band shift and a spectral weight transfer for Type-I and Type-II surfaces~\cite{lee2021distinguishing}, respectively. In contrast, subsequent scanning tunneling microscopy and spectroscopy (STM/STS) studies have shown a larger gap spectrum on the surface terminated with a single layer rather than a smaller gap~\cite{wu2022effect}. Moreover, the larger gap domain has been suggested to host a QSL phase, as evidenced by the band shift in a magnetic field~\cite{butler2023behavior, he2023spinon, he2023spinon2}. These results strongly imply that subsurface stacking configurations, which include stacking faults induced by undimerized layers, are crucial for reconciling the discrepancy in the measured tunneling spectra~\cite{lee2023charge}. However, experimental verification of subsurface stacking faults, their distribution and impact on the electronic structure of the surface remains limited.

In this work, we employ STM and STS measurements to visualize distinct insulating domains which form as a result of the change in stacking configurations across the surface of bulk 1\textit{T}-TaS${_2}$. Our observations reveal the coexistence of Type-I and Type-II domains near a monolayer step on the topmost layer~\cite{butler2020mottness, lee2021distinguishing}. Interestingly, in proximity to the monolayer step, both the upper and lower layers are predominantly occupied by the Type-I termination, suggesting that it is more energetically favorable for the stacking fault to reside beneath the surface. Furthermore, spatially resolved STS measurements over a large area reveal three distinct Type-I domains separated by both surface and subsurface domain walls. These distinct Type-I domains show a rigid band shift and band edge broadening, highlighting the interplay between subsurface stacking configurations and the electronic properties of the surface state. Finally, by combining our high-resolution STS data with density functional theory (DFT) calculations, we demonstrate how different subsurface stacking orders can lead to the observed electronic structures of the surface.

\bigskip
\noindent\textit{\bf Results and Discussion} 

\begin{figure*}
\includegraphics[width=\linewidth]{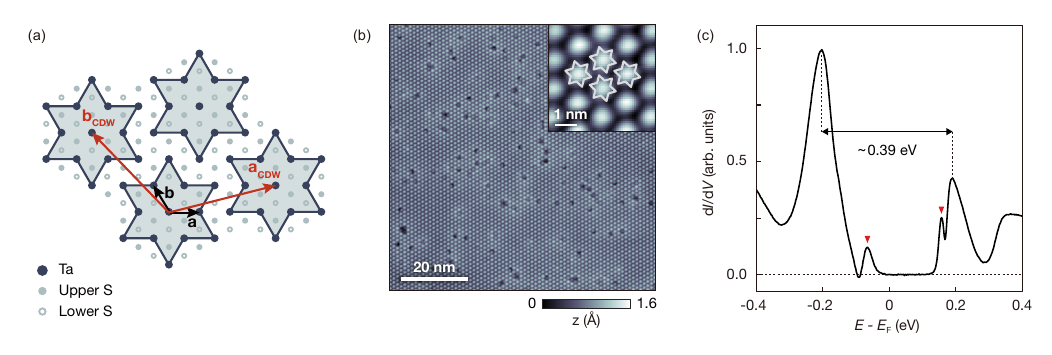}
\caption{Insulating commensurate CDW phase of 1\textit{T}-TaS${_2}$. 
a) A schematic representation of the atomic structure of 1\textit{T}-TaS${_2}$ with the Star of David (SD) superstructure. The black and red arrows correspond to the unit vectors of the $1\times1$ and the $\sqrt{13}\times\sqrt{13}$ lattices, respectively. 
b) An STM image of the commensurate CDW phase in 1\textit{T}-TaS${_2}$ ($V_{\rm set}=-400$ mV, $I_{\rm set}=50$ pA). Each bright protrusion represents an SD (inset). 
c) The representative $\mathrm{d}I/\mathrm{d}V$ spectrum acquired on the surface of a Type-I domain ($V_{\rm set}=-400$ mV, $I_{\rm set}=50$ pA, $V_{\rm mod}=5$ mV). The band gap, approximately $0.39$ eV, is marked by a black arrow. Our spectrum exhibits enhanced band edge features marked by red arrows.}\label{Figure1}
\end{figure*}

Our data are acquired in the commensurate CDW phase, characterized by the presence of a triangular SD superlattice (Figure~\ref{Figure1}a). A representative STM image (Figure~\ref{Figure1}b) shows the SD clusters as bright protrusions (inset). The dark spots in the STM image acquired at $-400$ mV are intrinsic defects related to sulfur (S) vacancies~\cite{cho2015interplay, lutsyk2023influence} whose features are more enhanced at the lower bias due to their in-gap states (Figure S1 in Supplementary Information). Figure~\ref{Figure1}c shows a representative differential conductance ($\mathrm{d}I/\mathrm{d}V$) spectrum acquired at a region away from intrinsic defects. The spectrum displays two prominent peaks located near $\pm 200$ meV, which show upward band bending near intrinsic defects or CDW domain walls. These insulating spectral features have been commonly observed in STS measurements~\cite{butler2020mottness, lee2021distinguishing, cho2017correlated}. Notably, our high-resolution STS spectra enable us to finely resolve the band edges as spectral peaks (see red arrows in Figure~\ref{Figure1}c) with negative differential resistance, which have usually been resolved as shoulders in the tunneling spectra. We attribute the peculiar characteristics to resonant tunneling between localized states of the sample and an atomically sharp STM tip~\cite{lyo1989negative}. These enhanced spectral features allow us to distinguish domains that have subtle differences in the edges of their electronic bands. The origin of the edges will be discussed in the subsequent section.

\begin{figure*}
\includegraphics[width=\linewidth]{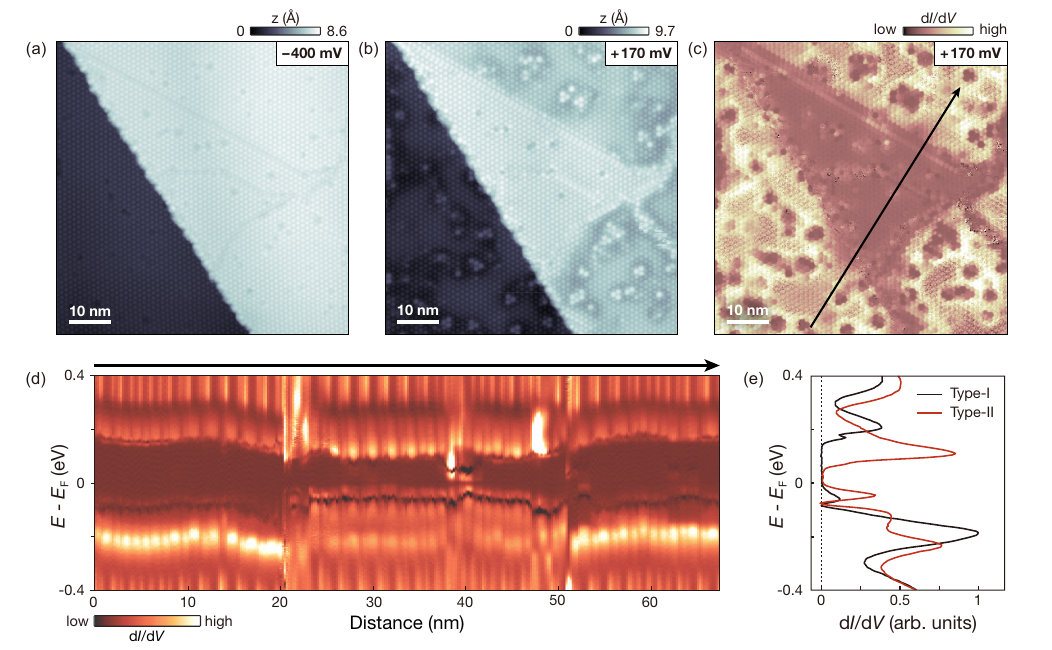}
\caption{Coexistence of dimerized (Type-I) and undimerized (Type-II) layer terminations on the upper terrace of a step edge.
a-b) STM images of the surface of bulk 1\textit{T}-TaS${_2}$ with the upper and lower terraces separated by a monolayer step obtained at different bias voltages (a $V_{\rm  set}=-400$ mV, $I_{\rm  set}=100$ pA and b $V_{\rm  set}=+170$ mV, $I_{\rm  set}=100$ pA, $V_{\rm  mod}=5$ mV). 
c) The $\mathrm{d}I/\mathrm{d}V$ map was acquired simultaneously with the STM image shown in b (See Methods). 
d) Line $\mathrm{d}I/\mathrm{d}V$ spectra acquired along the line marked by the arrow shown in c. 
e) Averaged $\mathrm{d}I/\mathrm{d}V$ spectra of the Type-I (black) and Type-II (red) domains acquired from defect-free areas ($V_{\rm set}=-400$ mV, $I_{\rm set}=200$ pA, $V_{\rm mod}=5$ mV).
}\label{Figure2}
\end{figure*}

We first present STM results acquired near a monolayer step to examine the electronic structure on both Type-I and Type-II surface terminations. The STM image (Figure~\ref{Figure2}a) shows bright and dark regions corresponding to the upper and lower terraces, respectively. A subsequent STM image at a reduced bias voltage of $+170$ mV (Figure~\ref{Figure2}b) displays a uniform electronic structure across the lower terrace, characterized by a consistent superstructure and absence of domain walls. In contrast, the upper terrace exhibits two types of domains with different electronic structures, separated by surface domain walls. The step height is consistent with the thickness of a single layer 1\textit{T}-TaS${_2}$, but is slightly higher at the step edge (Figure S2 in Supplementary Information). This implies that a higher conductance domain locally forms close to the step edge and rearranges the vertical charge order across the domain wall to maintain a similar electronic structure to the lower terrace~\cite{lee2023charge}.

The $\mathrm{d}I/\mathrm{d}V$ map (Figure~\ref{Figure2}c), simultaneously acquired with the STM image (Figure~\ref{Figure2}b), clearly illustrates the presence of two distinct types of domains with varying contrasts on the upper terrace: a dark triangular center domain and bright adjacent domains. Notably, the contrast of the bright domains on the upper terrace matches that of the lower terrace. A series of $\mathrm{d}I/\mathrm{d}V$ spectra (Figure~\ref{Figure2}d), taken along the black arrow in Figure~\ref{Figure2}c, indicates that the bright domains on both terraces indeed share identical electronic structures. Furthermore, it reveals that while both the bright and dark domains are insulating, their spectra exhibit significant differences in terms of bandgap values and peak intensities (Figure~\ref{Figure2}e): the peak-to-peak bandgaps for the two domain types are around $0.39$ eV and $0.2$ eV, respectively.  Since the contrast in the $\mathrm{d}I/\mathrm{d}V$ map highlights the relative difference in the electronic structures, the domain walls with in-gap states in larger gap domains are more prominent than those in smaller gap ones (See Experimental Section). A collection of $\mathrm{d}I/\mathrm{d}V$ maps obtained near the step edge (Figure S3 in Supplementary Information) demonstrates that the larger gap spectrum is dominant across the surface, with the smaller gap spectrum confined to small domains near the step edge.

According to previous DFT calculations~\cite{lee2023charge} and many-body theoretical calculations~\cite{petocchi2022mott}, the larger bandgap of $0.39$ eV results from interlayer hybridization in the dimerized layer (Type-I) and the smaller gap of $0.2$ eV arises from the correlated Mott gap in the undimerized surface layer (Type-II). The latter, a half-filled layer, exhibits correlated insulating behavior with strong on-site Coulomb repulsion, otherwise, it shows metallic behavior. This correlated insulating layer is expected to have a smaller gap than the band insulating layer with an appropriate Coulomb energy, which stabilizes the dimerized configuration. From the characteristic $\mathrm{d}I/\mathrm{d}V$ spectra of the two different domains (Figure~\ref{Figure2}e), we can identify the areas with higher and lower $\mathrm{d}I/\mathrm{d}V$ intensities in Figure~\ref{Figure2}c as Type-I and Type-II domains, respectively (See Experimental Section).

The crossover from Type-I to Type-II domains within a terrace is incompatible with the dimerization scenario for the insulating ground state in the layered CDW material. Prior STM studies have demonstrated that the undimerized layer is a Mott insulator with a smaller gap~\cite{butler2020mottness} and exhibits doping-induced spectral weight transfer instead of a rigid band shift~\cite{lee2021distinguishing}. However, subsequent STM results have shown surfaces with the larger gap spectrum, even when the terraces are separated by a monatomic step~\cite{wu2022effect, park2023stacking}. These observations have attracted renewed attention to the underlying physics of the ground state. Importantly, it was predicted that the undimerized layer (Type-II) is less stable than the dimerized layer at the surface, leading to the surface reconstruction of the CDWs~\cite{lee2023charge}. In systems with undimerized layers at the surface, CDW phase shifts occur via the lateral bond-relaxation of the SD clusters in the second layers. This reconstruction relocates the undimerized layer from the top to the third layer, leaving a dimerized layer on the surface (Type-I with a subsurface fault). We can interpret this unique reconstruction as the vertical bond relaxation driven by the lateral bond relaxation. The unstable Type-II surface can only manifest in small domains near surface steps, where electronic structures at the domain boundaries can overcome the surface energy instability within the domain. Our observations, in which most of the surface is populated by the Type-I domains with the Type-II domains existing primarily near the step edge, fully corroborate these theoretical predictions (Figure S3 in Supplementary Information). It is worth noting that our samples were cleaved at room temperature and subsequently cooled to cryogenic temperatures, ensuring that the vertical configuration of CDW stacking is stabilized during the commensurate CDW formation around 180 K. This strongly implies the presence of buried undimerized layers, which have been identified, within the otherwise fully dimerized ground state configuration.  

\begin{figure*}
\includegraphics[width=\linewidth]{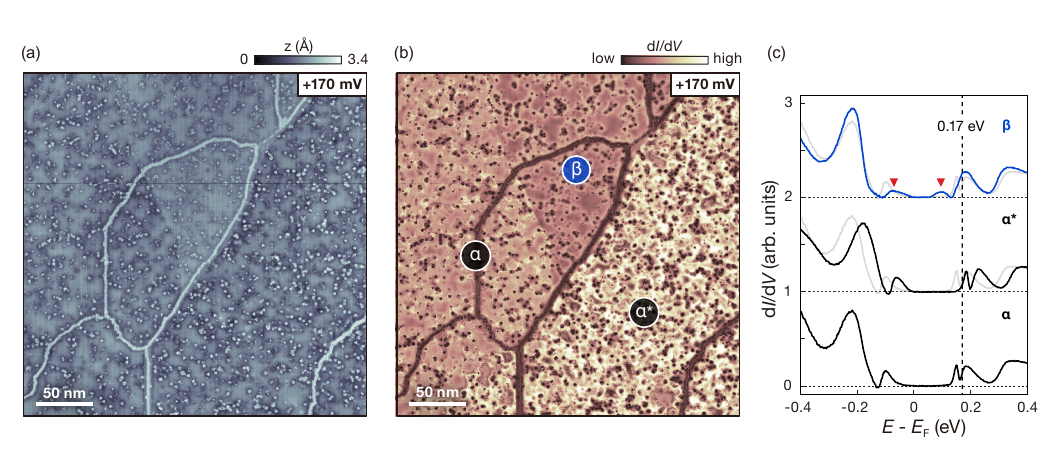}
\caption{Three different Type-I insulator domains on a flat surface.
a-b) Three different domains shown by both the STM image (a) and the $\mathrm{d}I/\mathrm{d}V$ map (b) ($V_{\rm set}=+170$ mV, $I_{\rm set}=50$ pA, $V_{\rm mod}=5$ mV).
c) $\mathrm{d}I/\mathrm{d}V$ spectra of the three domains ($\alpha$, $\alpha^{\rm *}$ and $\beta$) indicated in b averaged over areas not affected by local defects ($V_{\rm set}=-400$ mV, $I_{\rm set}=50$ pA, $V_{\rm mod}=5$ mV). Each spectrum is vertically shifted with equal spacing for clarity. The vertical dashed line indicates the energy for the maps shown in (a) and (b). The higher $\mathrm{d}I/\mathrm{d}V$ intensity in the tunneling spectra corresponds to the lower contrast in the $\mathrm{d}I/\mathrm{d}V$ map (b), which is simultaneously acquired with the topographic image (a) (see Experimental Section) For comparison, the spectrum of domain-$\alpha$ is superimposed as a faint gray line. There is an upward band shift for domain-$\alpha^{\rm *}$ and a band edge broadening (marked by red arrows) for domain-$\beta$.
}\label{Figure3} 
\end{figure*}

Although the topmost layer is likely to be dimerized, there can be various deviations from the ground state, forming heterogeneous domain configurations~\cite{lee2023charge,ishiguro1991electron}. To demonstrate the presence of distinct stacking orders, we visualize the electronic structure of a large flat terrace without a step edge (Figure~\ref{Figure3}a). By selecting a bias voltage close to the band edge, we highlight the spatial variation of the band edge that can be influenced by subsurface stacking orders (Figure S4 in Supplementary Information). Remarkably, the $\mathrm{d}I/\mathrm{d}V$ map of the same area (Figure~\ref{Figure3}b) shows three distinct insulating domains (marked by $\alpha$, $\alpha^{\rm *}$ and $\beta$) with varying intensities, which up to now have not been thoroughly investigated. The bright (dark) lines in Figure~\ref{Figure3}a (Figure~\ref{Figure3}b) represent surface domain walls resulting from CDW phase shifts (Figure S5 in Supplementary Information). Additionally, the consistency of our tunneling spectra with previous studies~\cite{lee2019origin,butler2020mottness,lee2021distinguishing}, which indicate dimerization of the first and second layers, leads us to attribute the sharp interfaces lacking CDW phase shifts to subsurface domain walls. These domain walls may exist in the third layer or below, giving rise to domains with different stacking orders. Therefore, these domain configurations and their interfaces strongly support the presence of distinct stacking configurations beneath the surface, resulting in a heterogeneous surface electronic structure.

Figure~\ref{Figure3}c shows the representative $\mathrm{d}I/\mathrm{d}V$ spectra of the three distinct insulating domains indicated in Figure~\ref{Figure3}b. They exhibit the characteristic Type-I spectrum, implied by the large band gap of about 0.39 eV~\cite{lee2019origin, petocchi2022mott, butler2020mottness}. Despite each having dimerized layers at the surface, there are some clear differences: First, domain-$\alpha^{\rm *}$ shows a notable rigid band shift of about $45$ meV compared to domain-$\alpha$ without a significant change in the overall spectral shape. Second, domain-$\beta$ exhibits broader band edges (marked by red arrows in Figure~\ref{Figure3}c) than domain-$\alpha$, while the two prominent peaks seem identical. It is worth noting that the variations in the surface electronic structure are not local features caused by defects or domain walls (Figure S6 in Supplementary Information), as evidenced by the homogeneous background $\mathrm{d}I/\mathrm{d}V$ contrast across each domain in Figure~\ref{Figure3}b.

\begin{figure*}
\includegraphics[width=\linewidth]{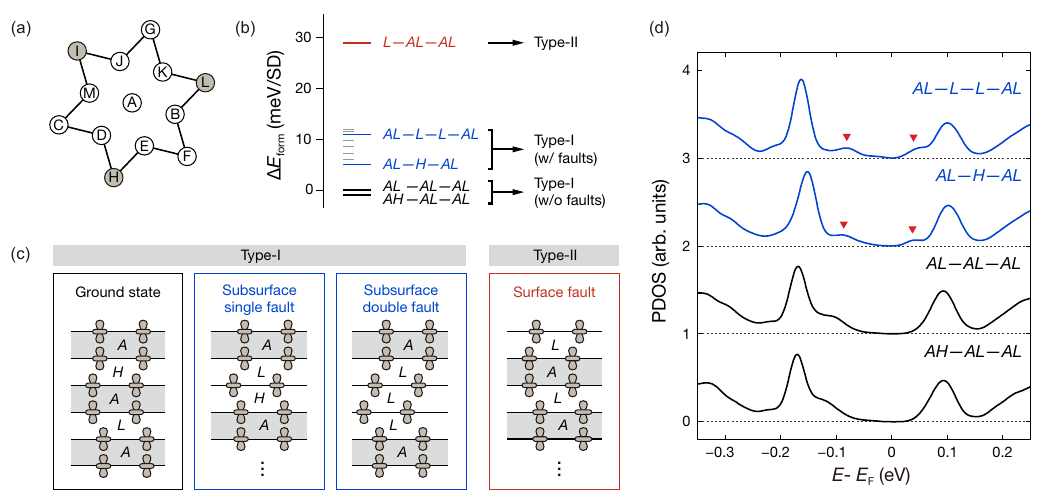}
\caption{Formation energies and electronic structures of the various stacking configurations.
a) A schematic representation of the SD supercell with 13 Ta atoms, each labelled in line with previous studies which have identified \textit{L}, \textit{I} and \textit{H} (identical by three-fold rotational symmetry, shaded in gray) as the most stable sliding configurations among the 12 possibilities (B-M)~\cite{lee2019origin}.
b) Direct comparison of the formation energies of two different types of stacking configurations and their variations: fully dimerized (black), subsurface undimerized (blue) and surface undimerized configurations (red). Formation energies of other possible stacking configurations, marked by short gray lines, and their corresponding PDOS spectra are presented in Figure S7.
c) A visual representation of four distinct stacking configurations discussed in (b). Each object denotes the position of the supercell center on individual layers, with the shaded regions (gray) indicating dimerization between adjacent layers.
d) The calculated PDOS spectra of the four different Type-I stacking configurations identified in (b). Each spectrum is vertically shifted with equal spacing, in order of their formation energies, with the lowest at the bottom. The red arrows point to the wider band edges caused by undimerized subsurface layers.}\label{Figure4}
\end{figure*}

Among the three different Type-I domains, we observed that domain-$\alpha$ and domain-$\alpha^{\rm *}$ appear most frequently on multiple cleaved surfaces. This implies that domain-$\alpha$ (or domain-$\alpha^{\rm *}$) has a lower formation energy compared to domain-$\beta$. To identify the stacking configurations, we adopt the notations from a previous work~\cite{lee2019origin} (Figure~\ref{Figure4}a). Our theoretical calculations clearly show that there are distinct Type-I configurations with energies lower than the Type-II (colored red in Figures~\ref{Figure4}b and c). Note that the layered material likely has all-dimerized configurations with diverse stacking orders (colored black in Figures~\ref{Figure4}b and c). In addition, one or more undimerized layers can be found in the bulk~\cite{ishiguro1991electron}, leading to a slightly higher formation energy (colored blue in Figures~\ref{Figure4}b and c). A comprehensive set of calculated formation energies for various stacking configurations is provided in Figure S7 in Supplementary Information. Although we cannot precisely determine the number of undimerized layers in the bulk sample, all-dimerized configurations are likely connected to domain-$\alpha$ and domain-$\alpha^{\rm *}$, while the other Type-I configurations are associated with domain-$\beta$, with subsurface faults within a few layers close to the surface.  

We now focus on the two distinct tunneling spectroscopic features of the surface density of states (DOS) of domain-$\alpha^{\rm *}$ and domain-$\beta$ with respect to domain-$\alpha$. The most obvious difference is that domain-$\beta$ has broader band edges compared to domain-$\alpha$ (Figure~\ref{Figure3}c). Our calculated projected density of states (PDOS) spectra reveal that the spectral features are related to the existence of the undimerized layer underneath the surface (red arrows on blue curves in Figure~\ref{Figure4}d).  The layer-selective PDOS spectra (See Figure S7 in in Supplementary Information) clearly show that the minor peaks at the band edges in the top-layer PDOS can be attributed to the coupling with the principal peaks of the third-layer central Ta atoms. The band gap for the subsurface dimerized layer is estimated to be 0.18 eV. In contrast, the band gap of the subsurface undimerized layer is around 0.11 eV. These results agree with the experimentally observed changes in the gap values of 0.25 eV and 0.16 eV determined by the minor peaks in domain-$\alpha$ and domain-$\beta$, respectively.  

On the other hand, domain-$\alpha^{\rm *}$ exhibits spectral characteristics identical to domain-$\alpha$ with an upward band shift (Figure~\ref{Figure3}c). These features are well consistent with the calculated spectra for all-dimerized configurations (black curves in Figure~\ref{Figure4}c). The band shift is attributed to a higher defect density in domain-$\alpha^{\rm *}$ compared to domain-$\alpha$ (Figure S8 in Supplementary Information). The upward band bending near the intrinsic defects is a typical screening behavior against acceptor-type dopants in the insulating material~\cite{lutsyk2023influence}. Taken together, we conclude that domains with distinct all-dimerized configurations exist and while their defect densities are not crucial in determining their stacking orders, they do influence their Fermi levels ($E_{\rm F}$).
  
Very recently, magnetic field-dependent STM experiments~\cite{butler2023behavior} revealed suggestive fingerprints of the QSL phase predicted to exist in 1\textit{T}-TaS${_2}$. Specifically, certain Type-I spectra exhibited a downward shift in the presence of an external magnetic field, consistent with the theoretical predictions for QSL behavior~\cite{he2023spinon, he2023spinon2}. However, two issues must be addressed to establish this phenomenon as direct evidence for the QSL phase. First, it conflicts with the conventional understanding that Type-I domains do not host localized spins due to interlayer hybridization. Second, some Type-I spectra do not exhibit the magnetic field response. Our data reconcile these disparate observations by demonstrating the likely presence of Mott insulating layers beneath band insulating surfaces. Thus, the two different behaviors of the Type-I spectra under a magnetic field can be attributed to subsurface stacking faults, emphasizing that Type-I domains cannot be characterized solely by a single electronic structure.

\bigskip
\noindent\textit{\bf Conclusion}

In conclusion, we have identified multiple distinct insulating domains in a wide field of view on the surface of bulk 1\textit{T}-TaS${_2}$ with our high-resolution STS measurements and DFT calculations. We find that the material prefers to maintain a dimerized surface via surface reconstruction and that the subsurface stacking configuration can be altered due to surface reconstruction. The spatial variations in the surface DOS, identified by the reduction of the CDW gap, broadening of the band edges and a rigid band shift, are revealed to be determined by the differences in stacking configurations and the defect concentration, respectively. These findings provide important insights into reconciling the discrepancies in spectroscopic measurements and identifying this layered CDW material as a platform for a QSL phase. Since our results are related to vertical charge orders, they are envisaged to be highly relevant to many other layered CDW systems, helping to resolve long-standing open issues about competing phases and ground state formation.

\newpage
\noindent\textit{\bf Experimental Section}

\noindent\textit{\bf Synthesis and STM Measurements}

1\textit{T}-TaS${_2}$ crystals (HQ graphene) were synthesized by chemical vapor transport (CVT) with iodine as a transport agent. The 1\textit{T}-phase was confirmed by low-temperature micro-Raman spectroscopy (Figure S9 in Supplementary Information). Measurements of STM and STS were conducted using a commercial low-temperature STM (UNISOKU) at a temperature of 4.2 K in an ultra-high vacuum environment of $1\times10^{\rm -10}$ Torr. Mechanically sharpened Pt / Ir $(90/10)$ wires were used for STM tips after characterizing them on a gold (Au) or lead (Pb) (111) surface.  Bulk 1\textit{T}-TaS${_2}$ samples were cleaved at room temperature and then transferred to the STM head, which had been cooled to 4.2 K. A standard lock-in technique was used for $\mathrm{d}I/\mathrm{d}V$ measurements, with an ac voltage (a voltage modulation amplitude of 5 mV and a frequency of 613 Hz) added to the dc sample bias. Occasionally, $\mathrm{d}I/\mathrm{d}V$ maps display a reverse contrast to a topographic image, which maps the height of the STM tip with feedback to keep the tunneling current constant. As the STM tip moves away from the surface due to the higher conductance, $\mathrm{d}I/\mathrm{d}V$ signal gets weaker when it is acquired simultaneously with a topographic image~\cite{wittneven1998scattering}.

\bigskip
\noindent\textit{\bf Density Functional Theory Calculations}

DFT investigations were carried out using the Vienna Ab Initio Simulation Package (VASP) ~\cite{VASP1, VASP2}. The computational approach involved the projector augmented wave method~\cite{PAW}, the generalized gradient approximation for the exchange-correlation potential ~\cite{PBE}, and the DFT~+~$U$ formalism as formulated by Dudarev et al.~\cite{Dudarev1998}, setting the Hubbard $U$ parameter to 1.25 eV for Ta $5d$ orbitals. Van der Waals forces were taken into account by applying the Tkatchenko-Scheffler method~\cite{vdW-TS}. Electronic wave functions were described with a plane-wave basis set, with cutoff energies of 323 eV for bulk and 259 eV for surface calculations. Reciprocal space integration was conducted using a \(4 \times 4 \times 8\) k-point mesh within the Brillouin zone of the \(\sqrt{13} \times \sqrt{13} \times 1\) supercell. For surface models, a periodic slab geometry with seven or eight layers of TaS$_2$ and a vacuum separation of 20 \AA\ was employed. The relaxation of all atomic positions was continued until the forces on each atom were below 0.01 eV/\AA.

\bigskip
\noindent\textit{\bf Raman Scattering Measurements}

Raman spectroscopic experiments were carried out using a $\lambda = 561$ nm laser (Oxxius) with the sample mounted onto the cold finger of a He-flow cryostat (Oxford MicroStat). The laser was focused onto the sample with a spot diameter of 2 $\mu$m and a laser power of less than 0.1 mW. Raman-scattered light was dispersed through a Princeton Instruments Trivista spectrometer onto a charge-coupled device detector (PyLoN eXcelon).



\newpage
\captionsetup{justification=Justified} 

\newpage

\newpage

\newpage

\clearpage
\bibliographystyle{apsrev4-2} 
\def\bibsection{\noindent{\textbf{References}}}
\bibliography{multiple_insulating_domains} 

\clearpage
\noindent\textit{\bf Supplementary Information}

\renewcommand{\figurename}{\textbf{{Supplementary Fig.}}}
\renewcommand{\thefigure}{\textbf{\arabic{figure}}}
\setcounter{figure}{0} 

\captionsetup{justification=Justified} 
\begin{figure*}[htb!]
\includegraphics[width=\textwidth]{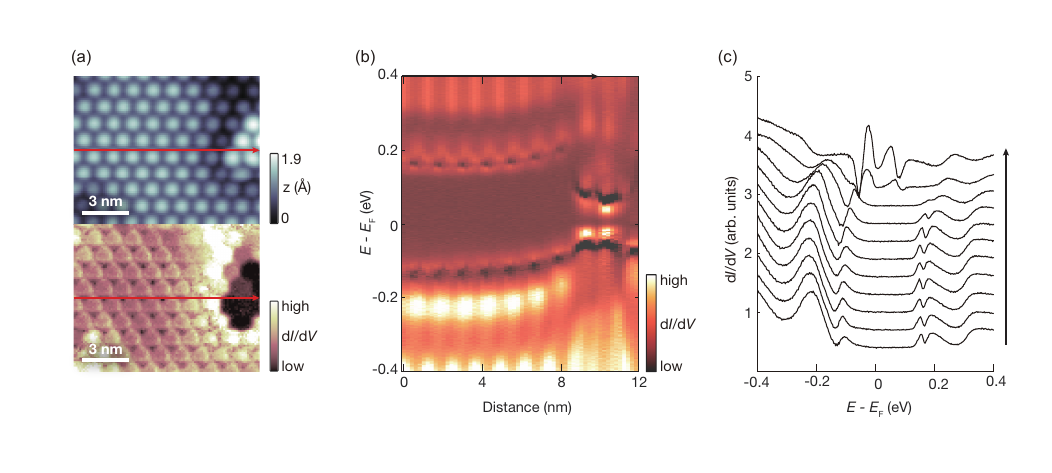}
\caption
{Defect spectra. a) STM image and $\mathrm{d}I/\mathrm{d}V$ map containing defects ($V_{\rm  set}=+200$ mV, $I_{\rm  set}=50$ pA). 
b) $\mathrm{d}I/\mathrm{d}V$ spectra taken across the red arrows in (a). 
c) A waterfall plot of the $\mathrm{d}I/\mathrm{d}V$ spectra along the black arrow in (b) with vertical shifts applied for clarity. Near the defects, the spectra show an upward shift in (b-c), which implies that these are acceptor-type defects. On top of the defects, in-gap states appear near the Fermi level ($E_{\rm F}$). 
}\label{Supplementary Fig.1} 
\end{figure*}

\newpage
\captionsetup{justification=Justified} 
\begin{figure*}[htb!]
\includegraphics[width=\textwidth]{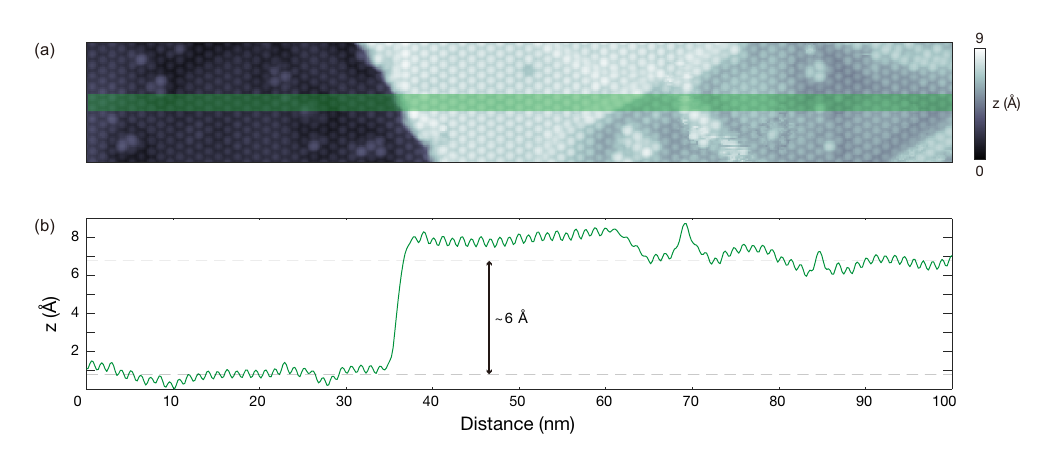}
\caption
{Height of the step edge in Figure 2 of the main text.
a) STM image acquired across the step edge shown in Figure 2 of the main manuscript ($V_{\rm  set}=+170$ mV, $I_{\rm  set}=100$ pA).
b) Averaged height profile across the area highlighted in green. A height of around 6 {\AA} is consistent with previous reports of a monolayer step~\cite{butler2020mottness,lee2021distinguishing}. Type-II surface is slightly higher than the thickness of the monolayer. It undergoes the charge rearrangement accompanied by the surface domain wall (marked by a red arrow), termed surface reconstruction, to become Type-I. 
}\label{Supplementary Fig.2} 
\end{figure*}

\newpage
\captionsetup{justification=Justified} 
\begin{figure*}[htb!]
\includegraphics[width=\textwidth]{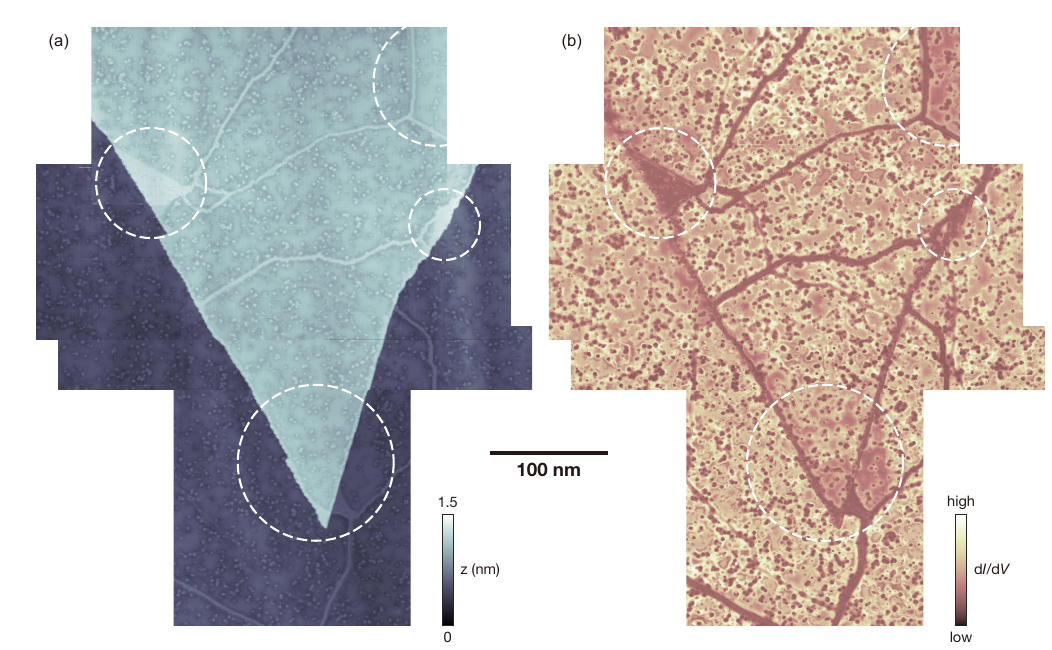}
\caption
{Large field of view of the various insulating states near a monolayer step edge in Figure 2.
a-b) Assembled a) STM images and b) $\mathrm{d}I/\mathrm{d}V$ maps show a larger field of view of the area shown in Figure 2 of the main manuscript. We identify various domains, including Type-II and Type-I* ($\beta$), denoted by white dotted circles. However, both the upper and lower terraces are dominated by Type-I domains, which can be seen by the overall homogeneous $\mathrm{d}I/\mathrm{d}V$ intensities ($V_{\rm  set}=+170$ mV, $I_{\rm  set}=21$ and $30$ pA).
}\label{Supplementary Fig.3} 
\end{figure*}

\newpage
\captionsetup{justification=Justified} 
\begin{figure*}[htb!]
\includegraphics[width=\textwidth]{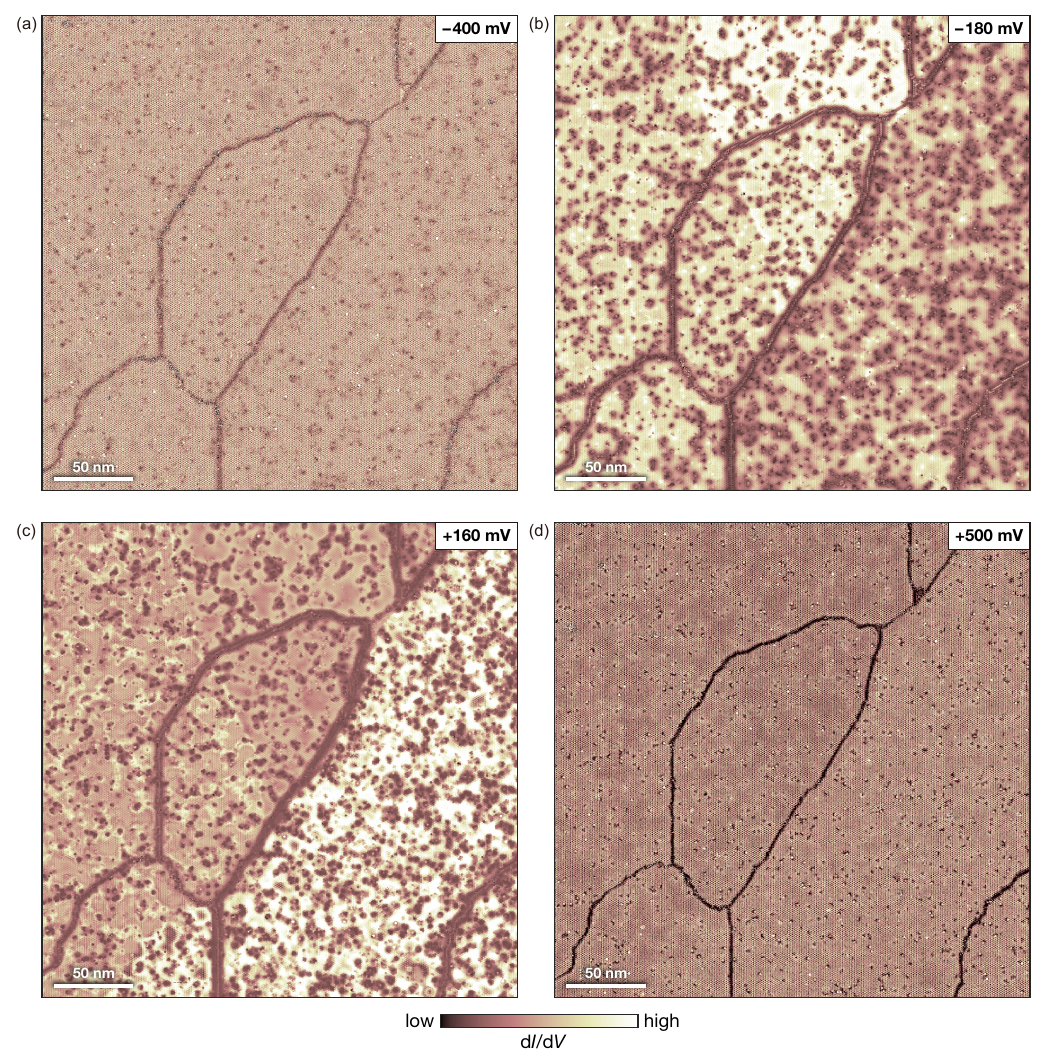}
\caption
{Series of $\mathrm{d}I/\mathrm{d}V$ maps at various bias voltages.
a-d) $\mathrm{d}I/\mathrm{d}V$ maps acquired in the same field of view as Figure 3 in the main text at various bias voltages  ($V_{\rm  set}=-400, -180, +160$ and $+500$ mV, $I_{\rm  set}=50$ pA). The boundaries dividing the domains become more pronounced with lower bias voltages (both positive and negative).
}\label{Supplementary Fig.4}
\end{figure*}

\newpage
\captionsetup{justification=Justified} 
\begin{figure*}[htb!]
\includegraphics[width=\textwidth]{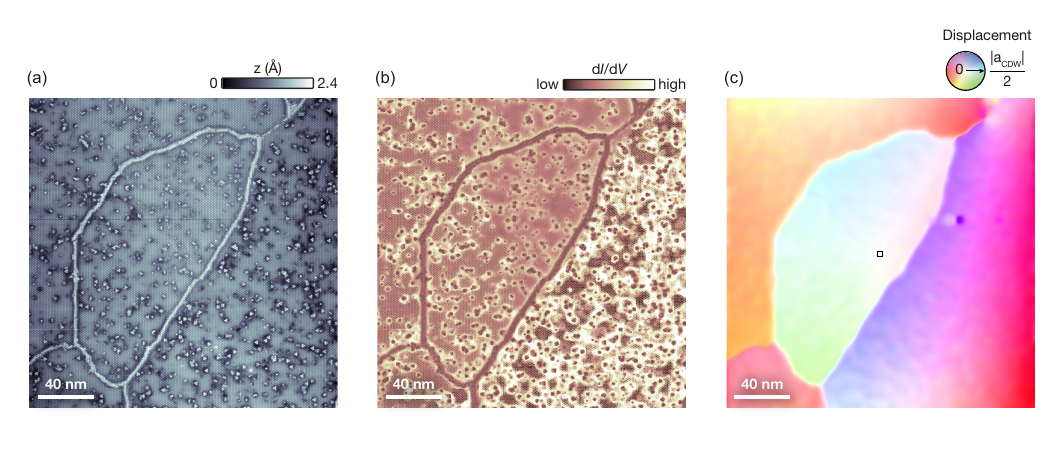}
\caption
{CDW displacement analysis.
a-b) Cropped areas of the STM image and $\mathrm{d}I/\mathrm{d}V$ map shown in Figure 3 of the main manuscript ($V_{\rm  set}=+170$ mV, $I_{\rm  set}=50$ pA).
c) CDW displacement map of the same area. We apply the Lawler-Fujita algorithm to analyze the CDW phases across the domain walls, with the black box at the center taken as reference\cite{butler2020mottness,lawler2010intra}. We identify the presence of abrupt changes in the CDW phases in (c) as surface domain walls, which are seen as bright and dark lines in (a) and (b), respectively.
}\label{Supplementary Fig.5} 
\end{figure*}

\newpage
\captionsetup{justification=Justified} 
\begin{figure*}[htb!]
\includegraphics[width=\textwidth]{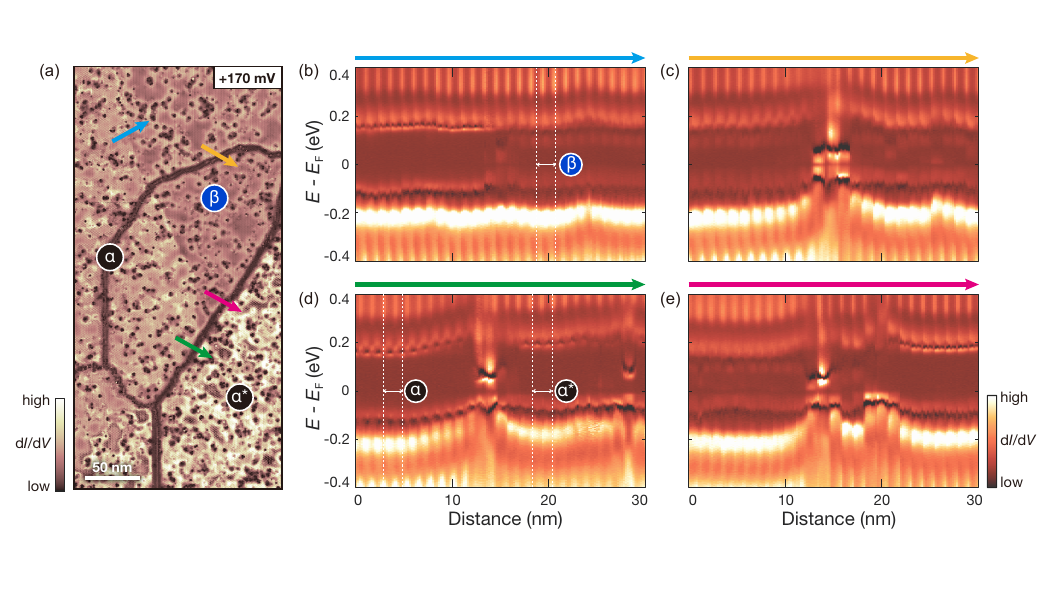}
\caption
{Multiple $\mathrm{d}I/\mathrm{d}V$ spectra taken across the surface and subsurface domain walls shown in Figure 3 in the main text.
a) Cropped $\mathrm{d}I/\mathrm{d}V$ map of Figure 3b.
b-e) $\mathrm{d}I/\mathrm{d}V$ spectra taken across b) a subsurface domain wall with different electronic structures (blue), c) a surface domain wall with the same electronic structure (yellow), d) a surface domain wall with different electronic structures (green) and e) a surface domain wall with the same electronic structure, continued by a subsurface domain wall with different electronic structures (magenta) ($V_{\rm  set}=-400$ mV, $I_{\rm  set}=200$ pA). The $\mathrm{d}I/\mathrm{d}V$ spectra for domains-$\alpha$, $\alpha$* and $\beta$ in Figure. 3e of the main text are obtained from regions unaffected by local defects and identified in b) and d) as white dotted lines. 
}\label{Supplementary Fig.6} 
\end{figure*}

\newpage
\captionsetup{justification=Justified} 
\begin{figure*}[htb!]
\includegraphics[width=\textwidth]{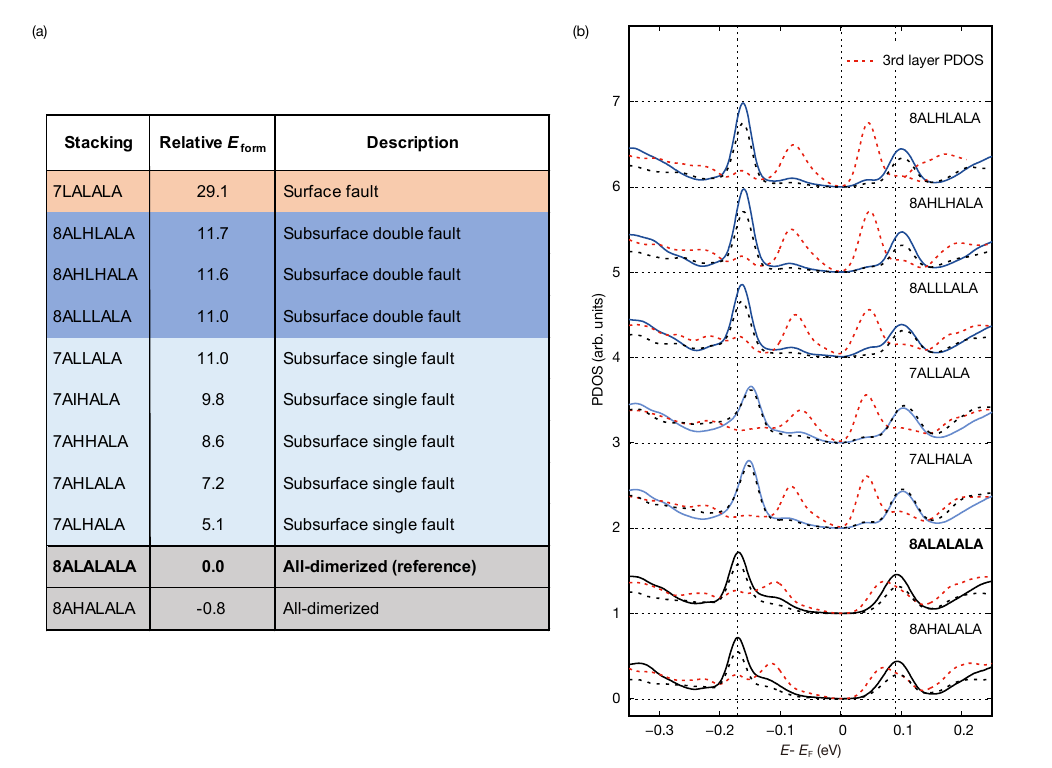}
\caption
{The formation energies of the favorable stacking configurations and their corresponding surface electronic structures. a) Relative formation energies of various stacking configurations, including all-dimerized layers (gray), subsurface single fault (bright blue), subsurface double fault (dark blue) and surface fault (red) with the 8ALALALA stacking configuration taken as the reference point. b) The PDOS spectra of distinct stacking configurations with the layer selective PDOS for the second (black dashed curves) and third layers (red dashed curves). The solid curves correspond to the surface PDOS acquired by integrating $d_{z^2}$ orbital contributions from 13 Ta atoms in the SD cluster.}\label{Supplementary Fig.7} 
\end{figure*}

\newpage
\captionsetup{justification=Justified} 
\begin{figure*}[htb!]
\includegraphics[width=\textwidth]{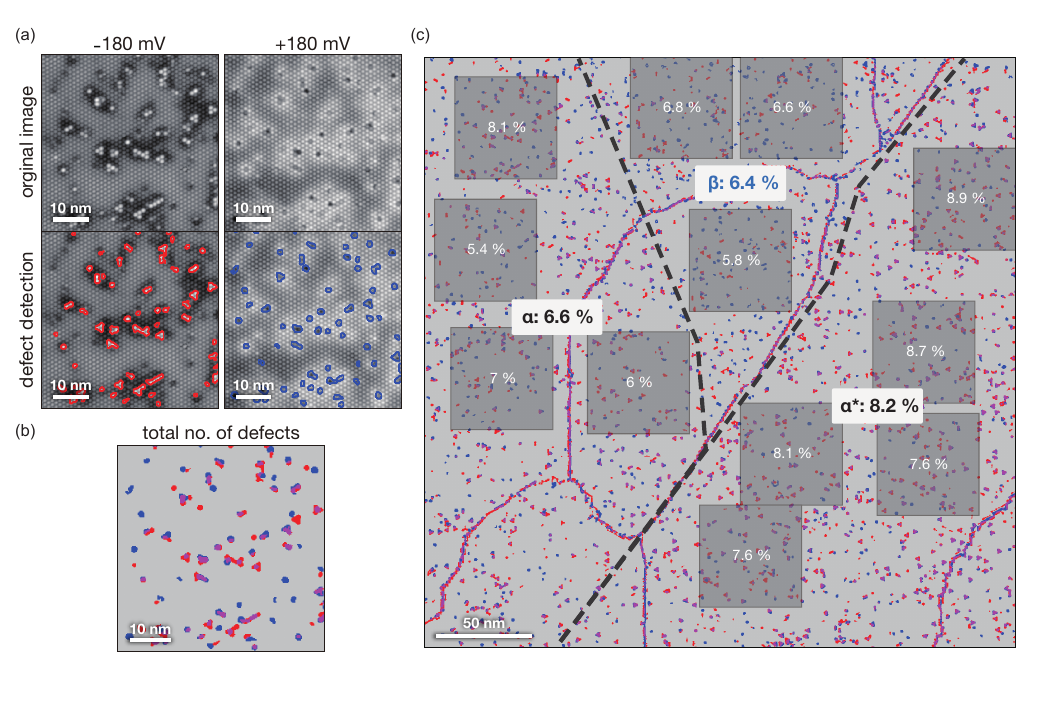}
\caption
{Distribution of defects in each domain.
a) STM images extracted from the same location as Figure 3(a-b) at negative and positive bias voltages ($V_{\rm  set}=-180$ and $+180$ mV). Defects are highlighted in red and blue for negative and positive bias voltages. At negative bias voltage, defects are usually shown as protrusions in the STM images, whereas at positive bias voltage, they are displayed as depressions. Defect sites in the STM images are identified and marked through image processing techniques.
b) Superposition of (red and blue) defects from (a). Some defects are shown only at negative or positive, which are identified as isolated blue or red dots. However, numerous defects are observable at both negative and positive bias voltages, which are shown as purple dots. We quantify the area covered by red, blue and purple to identify defect sites.
c) In the individual gray boxes, we note the ratio of defects, while the total averaged values of each domain are shown in the white boxes. Domain-$\alpha$* exhibits approximately 1.24 times more defects than the other domains.
}\label{Supplementary Fig.8} 
\end{figure*}

\newpage
Temperature- and position-dependent Raman measurements confirm the global crystalline structure of the sample. Indeed, both the thermal evolution of Raman-active modes through the different CDW phases, as well as Raman spectra taken at ten random spots across the mm-sized sample are in full accordance with the 1\textit{T} phase and clearly distinct from the 4$H$ one~\cite{nakashizu1984raman}. We note that with a laser spot diameter of 2 $\mu$m each spectrum contains rather mesoscopic information compared to the truly microscopic nature of the STM study. Interestingly, as can be seen in the color contour plot of panel b), we uncover subtle variations in the form of minor phonon shifts depending on the measurement position, which are greater than the spectral resolution limit of our experiment. While the origin of these variations is currently not fully understood, we interpret them as fingerprints of different Type-I : Type-II ratios, which are expected to slightly modify the energy of CDW-induced zone-folded phonons as well as CDW amplitude modes. We envision that future studies focusing on a detailed Raman mapping through various thermal cycling processes can pinpoint the origin and thereby directly link mesocopic Raman results to microscopic STM data.
\captionsetup{justification=Justified} 
\begin{figure*}
\includegraphics[width=\textwidth]{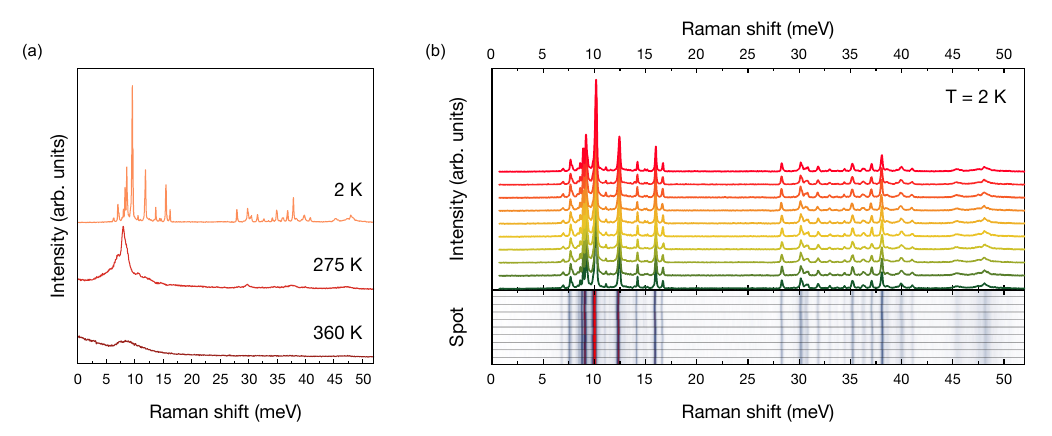}
\caption
{Raman spectroscopy data.
a) Raman spectra of 1$T$-TaS${_2}$, recorded in the ICCDW phase (360 K), the NCCDW phase (275 K) and the CCDW phase (2 K). 
b) Raman spectra of 1$T$-TaS${_2}$ in the CCDW phase at $T = 2$ K measured at ten random spots across the mm-sized sample.
}\label{Supplementary Fig.9} 
\end{figure*}

\end{document}